\documentclass[ajp,amsmath,amssymb,twocolumn]{revtex4-1} 
\usepackage[utf8]{inputenc}
\usepackage{amsmath}  
\usepackage{amsfonts} 

\usepackage[pdftex]{graphicx}

\usepackage{epstopdf}
\usepackage{siunitx}
\usepackage[pdftex=true,colorlinks=true,linkcolor=blue,citecolor=blue]{hyperref}

\makeatletter
\newlength \figwidth
\if@twocolumn
\else
\fi
\makeatother

\begin{document}


\title[Differential Dynamic Microscopy to characterize Brownian motion and bacteria motility]{Differential Dynamic Microscopy\\ to characterize Brownian motion and bacteria motility}

\author{David Germain}
\author{Mathieu Leocmach}
\author{Thomas Gibaud}
\email[Correspondence and requests for materials should be addressed to ]{thomas.gibaud@ens-lyon.fr}
\affiliation{Universit\'e de Lyon, Laboratoire de Physique, \'Ecole Normale Sup\'erieure de Lyon, CNRS UMR 5672, 46 All\'ee d'Italie, 69364 Lyon cedex 07, France}

\date{\today}

\begin{abstract}
We have developed a lab work module where we teach undergraduate students how to quantify the dynamics of a suspension of microscopic particles, measuring and analyzing the motion of those particles at the individual level or as a group. 
Differential Dynamic Microscopy (DDM) is a relatively recent technique that precisely does that and constitutes an alternative method to more classical techniques such as dynamics light scattering (DLS) or video particle tracking (VPT). DDM consists in imaging a particle dispersion with a standard light microscope and a camera. The image analysis requires the students to code and relies on digital Fourier transform to obtain the intermediate scattering function, an autocorrelation function that characterizes the dynamics of the dispersion. We first illustrate DDM on the textbook case of colloids where we measure the diffusion coefficient. Then we show that DDM is a pertinent tool to characterize biologic systems such as motile bacteria i.e.bacteria that can self propel, where we not only determine the diffusion coefficient but also the velocity and the fraction of motile bacteria. Finally, so that  our paper can be used as a tutorial to the DDM technique, we have joined to this article movies of the colloidal and bacterial suspensions and the DDM algorithm in both Matlab and Python to analyze the movies.
\end{abstract}

\maketitle

\section{Introduction}
\label{sec:level1}

Quantifying the dynamics of a suspension of microscopic particles consists in measuring and analyzing the motion of those particles at the individual level or as a group. Like hockey for Canadians or cricket for Indians and Pakistanis, quantifying the dynamics of a suspension of microscopic particles is the national sport of a large community of researchers in physics and biology. For example, a century ago, Perrin has characterized the motion of small particles in a liquid, an experiment that evidenced the Brownian motion and firmly proved the existence of atoms \cite{22_perrin2014atomes}. More recently, the motion of tracer particles has been used extensively in soft matter\cite{pt1990witten, pt1964reiner} to extract the mechanical properties such as viscosity or elasticity of fluids, gels\cite{Mason1997, Chen2010}, pastes, cell cytoplasm\cite{Fabry2001,Lau2003} and foods at scales unreachable by macroscopic techniques. In the past decade, the study of the collective motion of fish schools, bird flocks and bacteria swarms has lead to the emergence of a new field, active matter \cite{Bricard2013}.

Video Particle Tracking (VPT) and Dynamic Light Scattering (DLS) are two of the most well-known techniques to characterize the dynamics of a suspension of microscopic particles and that have been widely described in a teaching context. VPT consists in tracking the position of an individual particle as a function of time to digitalize its trajectory. It provides precise information on a limited region of interest of a sample \citep{Crocker1996,ajp2013catipovic,Maurer2014}. DLS consists in shining a laser through the particle suspension and monitoring the fluctuations of scattered intensity as a function of time\cite{ajp1969clark}. It yields average information about the dynamics of a sample \citep{18_sartor2003dynamic, ajp1999goldburg,ajp1969clark}.

Here, we present a tutorial for an alternative method called Differential Dynamic Microscopy (DDM) recently proposed by Cerbino and Trappe \cite{2_DDM}. DDM is intuitive as it deals with real-space video of the moving objects like VPT, however it uses digital Fourier transform to obtain the same kind of information as DLS.  Contrary to tracking algorithm, DDM algorithm is straightforward to implement at the programming level of most undergraduate students. We believe lab work on DDM is a great opportunity to become familiar with a microscope and with reciprocal units, the range of accessible wavenumbers and other nitty-gritty details necessary to tame the power of the Fourier transform\cite{ajp1976higgins, ajp2001whiford}. We have joined to this article movies of suspensions with different types of particles and the DDM codes in Matlab and Python so that the reader may reproduce the image analysis proper to DDM, see EPAPS.

In this article, we show how to apply DDM to two sets of different micrometer particles, colloids at first, then motile bacteria, i.e.bacteria that can self propel. In section \ref{sec:materials}, we present the materials and method we use to prepare the samples as well as the acquisition parameters. Section \ref{sec:ddm} presents the DDM principle and algorithm. In section \ref{sec:BrownSection}, we use DDM to characterize the Brownian motion of colloidal particles which has been widely studied in a teaching context~\cite{ajp1997lemons, ajp2006newburgh, ajp2006bergstein} and we determine the diffusion coefficient of the particles. In section \ref{sec:BactSection}, we apply DDM to \textit{salmonella} bacteria and show that it is possible to characterize entangled dynamics where the bacteria both diffuse and ``swim". In particular we determine the proportion of motile bacteria, their diffusion coefficient as well as their velocity. This last example highlight the impact of physical techniques on biology. Finally in section~\ref{sec:didac}, we focus of the didactic aspect of the lab work.

\section{Materials and methods}
\label{sec:materials}
\subsection{Colloidal particles}

We use polystyrene spheres (Density of polystyrene, $\rho_\text{c}$ = $\SI{1040}{\kilogram\per\meter\cubed}$) with a catalog radius of $R=\SI{0.50}{\micro\meter}$  (FluoSpheres\circledR{} F-13082 from Thermo Fisher Scientific) and a 10\% polydispersity. The commercial dispersion is diluted 50 times in deionized water (at $T=\SI{20}{\degree C}$ the viscosity of water is $\eta =\SI{1.002}{\milli\pascal\second}$, and its density is $\rho_\text{s}=\SI{1000}{\kilogram\per\meter\cubed}$) so that the distance between two individual particles is large compared to $R$, typically $20 R$. This concentration remains sufficiently high to observe enough particles in the camera field of view and to accumulate satisfactory statistics. Polystyrene refractive index is $\approx 1.6$, larger than the one of water $\approx 1.3$, so the particles are visible in bright field microscopy.

To observe the Brownian motion of particles in a Newtonian fluid, several conditions are required \citep{16_CollSusp}. The particle has to be colloidal, meaning that its size has to be far larger than the size of the solvent molecules, it is the case for the suspension we used (molecular diameter of water \citep{17_marcus1998properties}  $\SI{0.34}{\nano\meter}$). Furthermore, colloidal particles have to be in the dilute regime to avoid interaction between them. This is checked once the sample is made. As our particles do not have long distance interaction, we estimate that the sample is dilute when the mean distance between particles is higher than at least 10 $R$. Finally, the thermal agitation should be the physical process that dominates the dynamics of the colloidal particles. We ensure that the solvent is not flowing by using an immobile, sealed and thin optical cell with negligible temperature gradient. The sedimentation motion can be characterized by the P\'eclet number \citep{12_patankar1980numerical, ajp2009saka}: $\text{Pe}_\text{g} = \frac{E_\text{g}}{E_\text{th}}$, where $E_\text{g} = \frac{4 \pi R^3}{3}  (\rho_\text{c} - \rho_\text{s}) g  \times 2R$ is the variation of potential energy for a difference of altitude equal to the diameter $2R$ of the particle, $E_\text{th} = k_\text{B} T$ is the thermal energy, $\rho_\text{c}$ and $\rho_\text{s}$ the respective densities of the particle and the surrounding solution, $g$ the acceleration of gravity, $k_\text{B}$ the Boltzmann constant and $T$ the temperature. Here we have $\text{Pe}_\text{g} \approx 0.1$, indicating that the sedimentation can be neglected over Brownian motion. Equivalently in the time domain, our particle respectively sediments and diffuses on a distance equal to its own diameter on respective characteristic times $\tau_\text{s}=9\eta/(8(\rho_\text{c}-\rho_\text{s}) R)\approx\SI{100}{\second}$ and $\tau_\text{d}=24\pi\eta R^3/k_BT\approx\SI{2}{\second}$, so we have $\tau_\text{s}\gg\tau_\text{d}$. Absence of both flow and sedimentation will be verified \textit{a posteriori}. In presence of a flow the trajectory of the particle, that should be random, is biased in the direction of the flow.

\subsection{Bacteria}
In section \ref{sec:BactSection}, we study the motion of non-pathogenic bacteria \textit{Salmonella Typhimurium SJW1103}  (American Type Culture Collection, Manassas, VA, U.S.A.) \citep{21_fabrega2013salmonella}. This bacterium has the shape of a rod of length of $\SI{2}{\micro\meter}$ and diameter of $\SI{1}{\micro\meter}$ with flagella that insure self propulsion\citep{5_berg2000motile}. The global motion of this bacterium can be split into two modes~\cite{Nelson2004}. In the ``tumble'' mode, the motors rotate clockwise and independently, causing the bacterium to move erratically. In the ``run'' mode, the motors are synchronized and rotate counter-clockwise leading the bacterium to move ballistic and straight forward. As we gently mixed the culture medium during growth, the nutrient medium can be considered homogeneous in our samples and the bacteria are moving isotropically.

\textit{Salmonella Typhimurium SJW1103}  are stored in a freezer at $\SI{-80}{\degree C}$ in a mixture of water ($\simeq \SI{33}{\%} w$) and glycerol ($\simeq 66\% w$). First, using a sterile inoculation loop, we streak bacteria from the storage solution on a sterile agar/LB plates  ($\SI{500}{\milli\liter}$ of LB/agar was made of $\SI{5}{\gram}$ of NaCl, $\SI{5}{\gram}$ of Tryptone, $\SI{2.5}{\gram}$ of Yeast Extract, $\SI{7.5}{\gram}$ of Agar). Second, the agar plates are closed and placed in an incubator at $\SI{37}{\degree C}$. The agar plates are oriented such that the LB/agar gel is at the top, to prevent the condensation from disturbing the development of the bacteria. After $\simeq$ 12 hours, we observe the formation of monoclonal circular colonies. Third, Using an inoculation loop, we take a monoclonal colony of bacteria from LB/Agar plate and disperse it in a \textsc{Falcon} tube with $\simeq \SI{5}{\milli\liter}$ TN growth medium. The TN growth medium is sterile and composed of $\SI{4}{\gram\per\liter}$ of bacto-tryptone, $\SI{2.5}{\gram\per\liter}$ of NaCl and $0.4\% w$ of glycerol diluted in water. The \textsc{Falcon} tubes have an oxygen permeable cap which allows the bacteria to breath and limits evaporation. Those bacteria are pre-cultured for a night at $\SI{32}{\degree C}$ at a shaking speed of 300 rpm in an Incu-Shaker 10L. Finally, we collected $\SI{50}{\micro\liter}$ of the solution of bacteria with a sterile pipette and we put it in a new \textsc{Falcon} tube filled with $\simeq \SI{5}{\milli\liter}$ of TN growth medium. The tube is then placed in the Incu-Shaker 10L at 300 rpm and $\SI{32}{\degree C}$ for 1h30. Around this time the optical density (OD) at $\SI{600}{\nano\meter}$ is around OD=0.5 and most bacteria are ``swimming". It is important to grow bacteria in a nutrient poor media and to collect them early on, at low concentrations, otherwise the bacteria tend not to develop a flagella. More details about bacteria preparation can be obtained in ref\cite{Schwarz2015, ajp2010hagen}.

\subsection{Microscope slide}
\begin{figure}
\includegraphics[width=\figwidth]{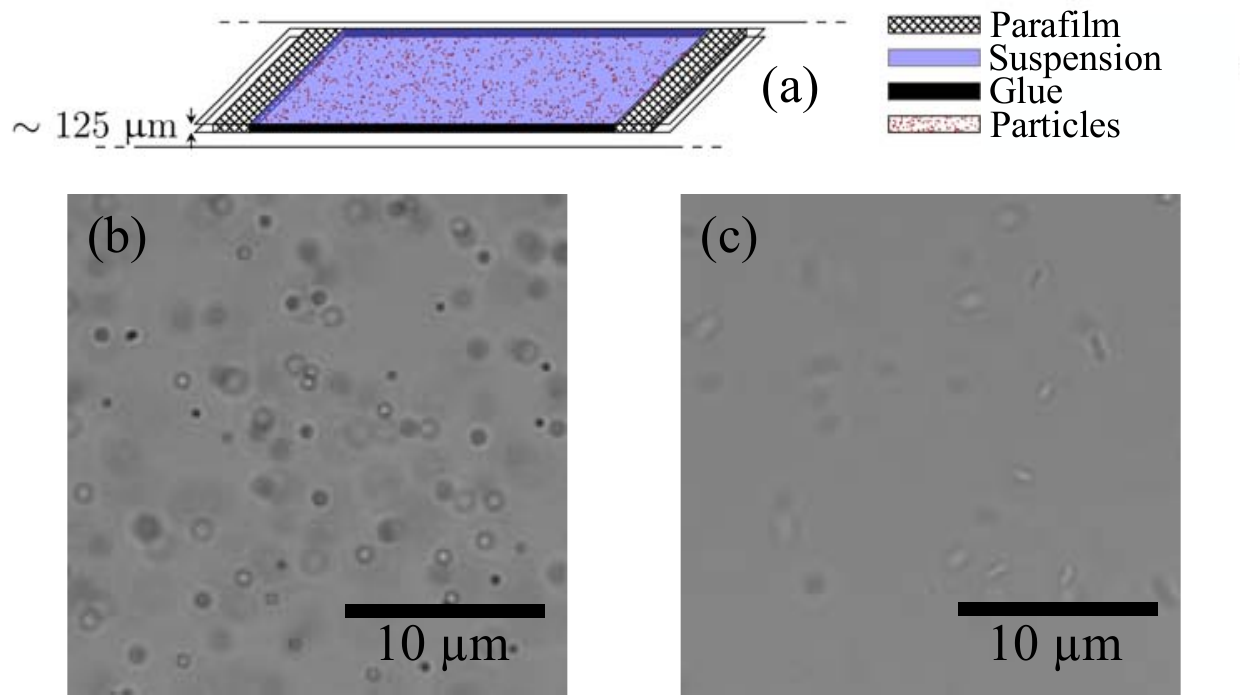}
\caption{Experimental setup. (a) Schematics of the microscope cell used to study the motion of particles, namely, colloids and bacteria. (b) Typical bright field image of our colloidal suspension. (c) idem with bacteria.}
\label{fig:Slide}
\end{figure}

For the microscope observations, the aqueous suspension of colloids or bacteria are enclosed in a home made hermetic optical cell, Fig.\ref{fig:Slide}.a and observed at room temperature, $T=\SI{20}{\degree C}$. The cell is composed of a  glass slide (\textsc{RS France}) and a cover slip (\textsc{Menszel-gl\"aser}) spaced by two stripes of paraffin film (\textsc{Bemis}) creating a gap between the glass slide and the cover slip of approximately $\SI{125}{\micro\meter}$. The optical cell is briefly heated on a hot plate so that the paraffin welds to the slide and the cover slip. The suspension of colloids or bacteria is then introduced into the slit by capillarity, and the slit is immediately sealed at both extremity using ultraviolet-cured glue (Norland Optical).

\subsection{Microscope and acquisition parameters}

The colloidal and bacterial suspensions are observed with light field microscope\cite{Mignard2015} (\textsc{nikon Eclipse Ti}) in bright field with an objective $10 \times$ of numerical aperture $N.A. = 0.3$, Fig.\ref{fig:Slide}.b-c. The focus is made in the middle of the microscope slide in $z$-direction so that we observe only particles able to move in the 3 dimensions. Images are acquired with a camera (\textsc{Hamamatsu ORCA-Flash 2.8}).

The images $I(\vec{r}, t)$ depends on the time $t$ when it was acquire. $\vec{r}$ is the coordinate of the pixels of the image.  $I(\vec{r}, t)$ is coded in $\SI{8}{bits}$  grey-scale: each pixel intensity is proportional to the incoming light intensity from the sample and varies linearly from  0 (black) to 255 (white). Using an exposure time of $\SI{1}{\milli\second}$, we adjusted the brightness of the microscope light in order to have a maximum number of pixel around a value of $\sim 120$ which minimize the amount of saturated pixels.

{The choice of the acquisitions parameters is a compromise between the spatial and the temporal resolution. With the $10\times$ objective, a pixel corresponds to $dL=\SI{0.645}{\micro\meter}$, a bit smaller than the optical resolution of the microscope, $\lambda/(2N.A.)\approx\SI{0.8}{\micro\meter} $. Bacteria or colloidal particles are about $\SI{1}{\micro\meter}$ and therefore correspond to a few pixels on the camera. We chose a $(\SI{512}{px})^2$ field of view which is large enough to capture the motion of 100 of particles and small enough to reach high acquisition frequencies, up to $\SI{400}{\hertz}$. At $\SI{400}{\hertz}$, it is impossible to directly send the data from the camera to the computer during the capture, so we have to temporarily save the acquisition on the buffer of the camera, and, once it is over, we send the data from the buffer of the camera to the computer. The camera buffer memory limits  to 4000 the number of images in one stack. To cover a wide range of time scales, we chose to acquire a first stack of $N$=4000 images at $\SI{400}{\hertz}$ and a second similar stack at $\SI{4}{\hertz}$. With this procedure, we cover time scales between $\SI{2.5e-3}{\second}$ and $\SI{1000}{\second}$ and length scales between $dL=\SI{0.645}{\micro\meter}$ and $L=\SI{330}{\micro\meter}$. 

$\hat{I}(\vec{q}, t)$ is the numerical 2D Fourier transform of $I(\vec{r}, t)$ obtained using Fast Fourier Transform (FFT) algorithm which is widespread and included in most high-level languages. $\vec{q}$ refers to the wave vector which norm is $q=2\pi/r$. $\hat{I}(q, t)$=$ \left\langle\hat{I}(\vec{q},t) \right\rangle_{\vec{q}}$ is the radial average of  $\hat{I}(\vec{q},t)$  obtained by averaging the value of all the pixels that are between $q$ and $q+dq$ from the center of this 2D spectrum. In Fourier space, the wave number increment, which also corresponds to the minimum wave number, is related to the image size, $dq=2\pi/L=\SI{0.019}{\micro\meter^{-1}}$. The maximum wave number is $q_{max}=2\pi/(2dL)=\SI{4.87}{\micro\meter^{-1}}$. Indeed according to the Nyquist–Shannon sampling theorem, the smallest wavelength measurable corresponds to a sinusoidal wave of period 2 pixels: 1 pixel for the positive part of sinusoidal wave and 1 pixel for the negative.

\section{Differential dynamic microscopy}
\label{sec:ddm}

\subsection{Autocorrelation function $f$}
\label{sec:ddm1}

Differential Dynamic Microscopy (DDM) aims at obtaining the auto-correlation function $f$\cite{ajp1995passmore} from a stack of images $I(\vec{r}, t)$. $f$ is sensitive to the dynamics of the system and measures the similarity of the statistical properties of images separated by a lag time $\Delta t$. In Fourier space, the expression of $f$ is simple,  thanks to the Wiener-Khintchine theorem. It is the product between the Fourier transform conjugate of the images at time $t$,  $\hat{I}^*(\vec{q}, t)$, and the Fourier transform of the image at time $t+\Delta t$, $\hat{I}(\vec{q}, t+\Delta t)$ normalized by the square norm of $\hat{I}(\vec{q}, t)$:
\begin{equation}
f(q, \Delta t) =\frac{\left\langle \hat{I}^*(\vec{q}, t) \hat{I}(\vec{q}, t+\Delta t) \right\rangle_{t, \vec{q}}}{\left\langle | \widehat{I}(\vec{q},t)|^2  \right\rangle_{t, \vec{q}} }
\label{eq:f}
\end{equation}
As the colloidal or the bacteria display stationary ergodic and  isotropic dynamics, we average over the initial time $t$ and the orientation of the wavevector $\vec{q}$ as symbolized by the brackets, so that $f$ no longer depends on $t$ and $\vec{q}$  but only on the lag time $\Delta t$ and the wave number $q$ respectively. 

For such systems, $f$ shows interesting general properties. $f$ decays with $\Delta t$ from 1 when the system has not changed ($\Delta t=0$) to 0 when the system has changed completely $(\Delta t\rightarrow\infty)$. The decay of $f(q, \Delta t)$ and its characteristic time $\tau(q)$ depend on the length scale via $q$. $\tau(q)$ is longer for large length scales, i.e. smaller $q$ because larger structures need longer time to decorrelate. The combined dependence of $f$ with $q$ and $\Delta t$ contains information about the physical origin of the decorrelation process. Based on dimension analysis, a diffusion process is characterized by its diffusion coefficient $D$ which scale as $[D]=[\text{length}]^2/[\text{time}]$ so that $f$ collapses on a master curve when plotted as a function of $q^2\Delta t$ (dimension of the inverse of the diffusion coefficient). In contrast, a ballistic process is characterized by its velocity $v$ which scale as $[v]=[\text{length}]/[\text{time}]$ and  $f$ collapses on a master curve when plotted as a function of $q\Delta t$ (dimension of the inverse of the velocity).

\subsection{DDM principle}

In DDM, $f$ is obtained as follow. First, we consider the difference between two images separated by $\Delta t$,

\begin{equation}
\Delta I(\vec{r},t, \Delta t) = I(\vec{r}, t+\Delta t) - I(\vec{r}, t),
\label{eq:DI}
\end{equation}

The subtraction remove all static artifacts such as dirt on the slide or on the microscope lenses. Second, we compute the DDM matrix $\mathcal{D}$: the squared normed  ensemble averaged on $t$ of the numerical 2D spatial Fourier transform $\widehat{\Delta I}$. The calculation of $\mathcal{D}$ allows us to recover $f$ as defined in Eq.\eqref{eq:f} thank to cross product term:

\begin{align*}
\mathcal{D}(\vec{q},\Delta t) \equiv& \left\langle \left|\widehat{\Delta I}\right|^2 \right\rangle_t\\
 =& \left\langle \left|\widehat{I}(\vec{q},t+\Delta t) - \widehat{I}(\vec{q},t)\right|^2 \right\rangle_t \\
=& \left\langle \left|\widehat{I}(\vec{q},t+ \Delta t)\right|^2 + \left|\widehat{I}(\vec{q},t)\right|^2\right. \\
&\left.- 2 \, \widehat{I}^*(\vec{q},t) \widehat{I}(\vec{q},t+\Delta t) \right\rangle_t\\
=& \underbrace{2\left\langle | \widehat{I}(\vec{q},t)|^2  \right\rangle_t}_{\equiv A(\vec{q})} \bigg[ 1-\underbrace{\frac{\left\langle \hat{I}^*(\vec{q}, t) \hat{I}(\vec{q}, t+\Delta t) \right\rangle_t}{\left\langle | \widehat{I}(\vec{q},t)|^2  \right\rangle_t }}_{f(\vec{q}, \Delta t)} \bigg]
\addtocounter{equation}{1}\tag{\theequation} \label{eq:DDM}
\end{align*}
}

Third, we radial average $\mathcal{D}(\vec{q},\Delta t)$ and therefore drop the dependence on the orientation of the wavevector $\vec{q}$. The contribution of the dark, shot and read-out noise\cite{Mignard2015,Joubert2011} of the camera add some noise to each images and is taken into account by adding a supplementary term $B(q)$. $B(q)$ is decorrelated  at all time and therefore independent of $\Delta t$.

\begin{equation}
\mathcal{D}(q,\Delta t) = A(q)\left[1-f(q, \Delta t)\right] + B(q)
\label{eq:D2f}
\end{equation}

The DDM matrix does not directly yield $f$. Ones need to correctly evaluate $A(q)$ and $B(q)$ to get $f$. Two different strategy can be adopted. Strategy 1: $A(q)$ and $B(q)$ are measured independently based on the properties of $\mathcal{D}$: at short times $f$=1, $\mathcal{D}(q,\Delta t\rightarrow 0) = B(q)$, and long times, $f$=0, $\mathcal{D}(q,\Delta t\rightarrow \infty) = A(q)+B(q)$. This first method give access directly to the autocorrelation function $f$ which can then be fitted. It is however necessary to measure small enough and  long enough lag time $\Delta t$ with respect to the derecorrelation time $\tau$ otherwise $B$ is overestimated and $A+B$ is underestimated, respectivel. Strategy 2 consists in fitting directly $\mathcal{D}$ with $A$ and $B$ as free parameters and a model for $f$. This last method is less demanding on the range of the lag time $\Delta t$ but it requires a model for $f$ and therefore prevent a scaling approach in the first place as described in section \ref{sec:ddm1}.

\subsection{DDM algorithm}
\label{sec:ddmalgo}

\begin{figure*}
	\includegraphics[width=\linewidth]{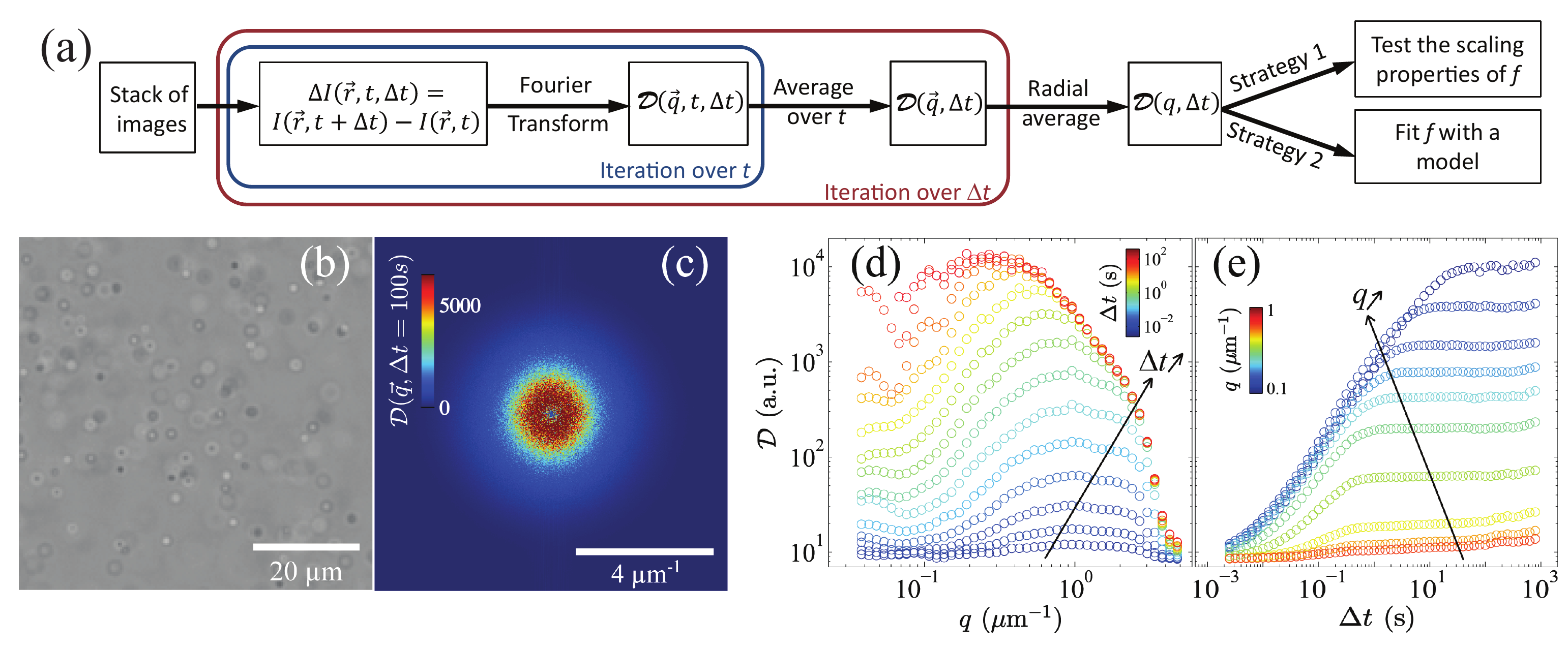}\\
	\caption{DDM principle. a) Schematic of the DDM algorithm. b) Image differences $\Delta I(\vec{r}, t,\Delta t=100 \text{ s})$. c) Square of the Fourier transform $|\widehat{\Delta I}|^2$ of the image in  (b). d) DDM matrix averaged and projected on $q$ for different lag times $\Delta t$. e) DDM matrix radial averaged and projected on $\Delta t$ for different $q$.  The results are obtained with the colloidal suspension.}
	\label{fig:D}
\end{figure*}

The DDM algorithm consists in two nested loops on $\Delta t$ and $t$ respectively, see Fig.~\ref{fig:D}a. At each iteration, we first open the couple of images $I(\vec{r},t)$ and $I(\vec{r},t+\Delta t)$, calculate $\Delta I(\vec{r},t, \Delta t)$ via Eq.~\eqref{eq:DI}, see Fig.~\ref{fig:D}b, and its Fourier spectrum $\left|\widehat{\Delta I}(\vec{r},t, \Delta t)\right|^2$, see Fig.~\ref{fig:D}c.  The loop on $t$ allows to compute at a fixed $\Delta t$ the time-averaged $\mathcal{D}(\vec{q}, \Delta t)$. This radial average yields $\mathcal{D}(q, \Delta t)$, see Fig.~\ref{fig:D}d. We then iterate on $\Delta t$.

A few tricks are performed to reduce calculation time. The role of the inner loop on $t$ is to gather statistics. At most it runs over $N-\Delta t$ couples of images which can be very expensive for short $\Delta t$. We found that limiting this number to 300 was enough provided that the initial times $t$ are evenly sampled across the accessible time window. Again to save calculation time, we logarithmically sampled $\Delta t$ with 10 points per decades which reduces the number of iteration of the outer loop from $N-1$=3999 to 10log$N$=35. With those optimizations, the calculation time falls to a few minutes. 

We run the DDM procedure on both stacks of images at $\SI{400}{\hertz}$ and $\SI{4}{\hertz}$ independently. We then merged the two sets of data by scaling the data at $\SI{4}{\hertz}$ so that both values at $\SI{0.25}{\second}$ are equal. We average the values of $\mathcal{D}$ at $\SI{4}{\hertz}$ and $\SI{400}{\hertz}$ in the overlap interval, from $\SI{0.25}{\second}$ to $\SI{10}{\second}$. We thus obtain $\mathcal{D} (q, \Delta t)$, see Fig.~\ref{fig:D}e, for $\Delta t$ from $\SI{2.5e-3}{\second}$ to $\SI{1000}{\second}$ a range of $\Delta t$ wide enough to correctly measure or fit $A$ and $B$.

The final step consist in analyzing $\mathcal{D}$ (Eq.~\ref{eq:D2f}) to extract information on dynamics of the observed dispersion. In the following, for both colloidal and bacteria suspensions, we decided to display $f$ rather than $\mathcal{D}$ because it is easier to interpret and to compare to DLS experiments which also yields $f$. We will use strategy 1 to apprehend the scaling properties of $f$ and strategy 2 to fit $f$ with the appropriate model. $\mathcal{D}$ is fitted in logarithmic scale in order not to attribute too much weight on points with high intensity and we dismiss timescales above \SI{200}{\second} where the statics is poor because $\mathcal{D}$ is averaged less than 4 times.

\section{Brownian motion and DDM}
\label{sec:BrownSection}

\subsection{The Brownian motion of colloids}
The Brownian motion is the unceasing and random motion of small particles suspended in a fluid at rest. It is due to the shocks between the solvent molecules and the colloidal particles~\cite{13_brown1828brief,9_einstein1906theory, 14_sutherland1905lxxv, 10_von1906kinetischen,ajp1997lemons,22_perrin2014atomes} and has often been reviewed in the litterature~\cite{ajp1997lemons,Nelson2004, ajp2006newburgh, ajp2006bergstein, Pearle2010}. For spherical Brownian particles  diffusing in the  background solvent, the autocorrelation function is\cite{18_sartor2003dynamic, ajp1999goldburg,ajp1969clark,8_berne2000dynamic}:
\begin{align}
f(q, \, \Delta t) &= \exp(-\Delta t/\tau_\text{d})
\label{eq:fdiff}\\
\text{with }\tau_\text{d} &= \frac{1}{Dq^2}.
\label{eq:relax}
\end{align}
$\tau_\text{d}$ is the characteristic diffusion time of the exponential decay and $D$ the diffusion coefficient of the particles. $D$ is increasing with the temperature $T$ and decreasing with the radius $R$ of the particle and the viscosity $\eta$ of the solvent according to the \textsc{Stokes-Einstein} formula \cite{Nelson2004, 16_CollSusp, ajp1997lemons, ajp2007jia}:
\begin{equation}
D = \frac{k_\text{B} T}{6 \pi \eta R}
\label{eq:se}
\end{equation}

\subsection{Results}
\label{sec:resultcolloid}
\begin{figure}
	\includegraphics[width=\figwidth]{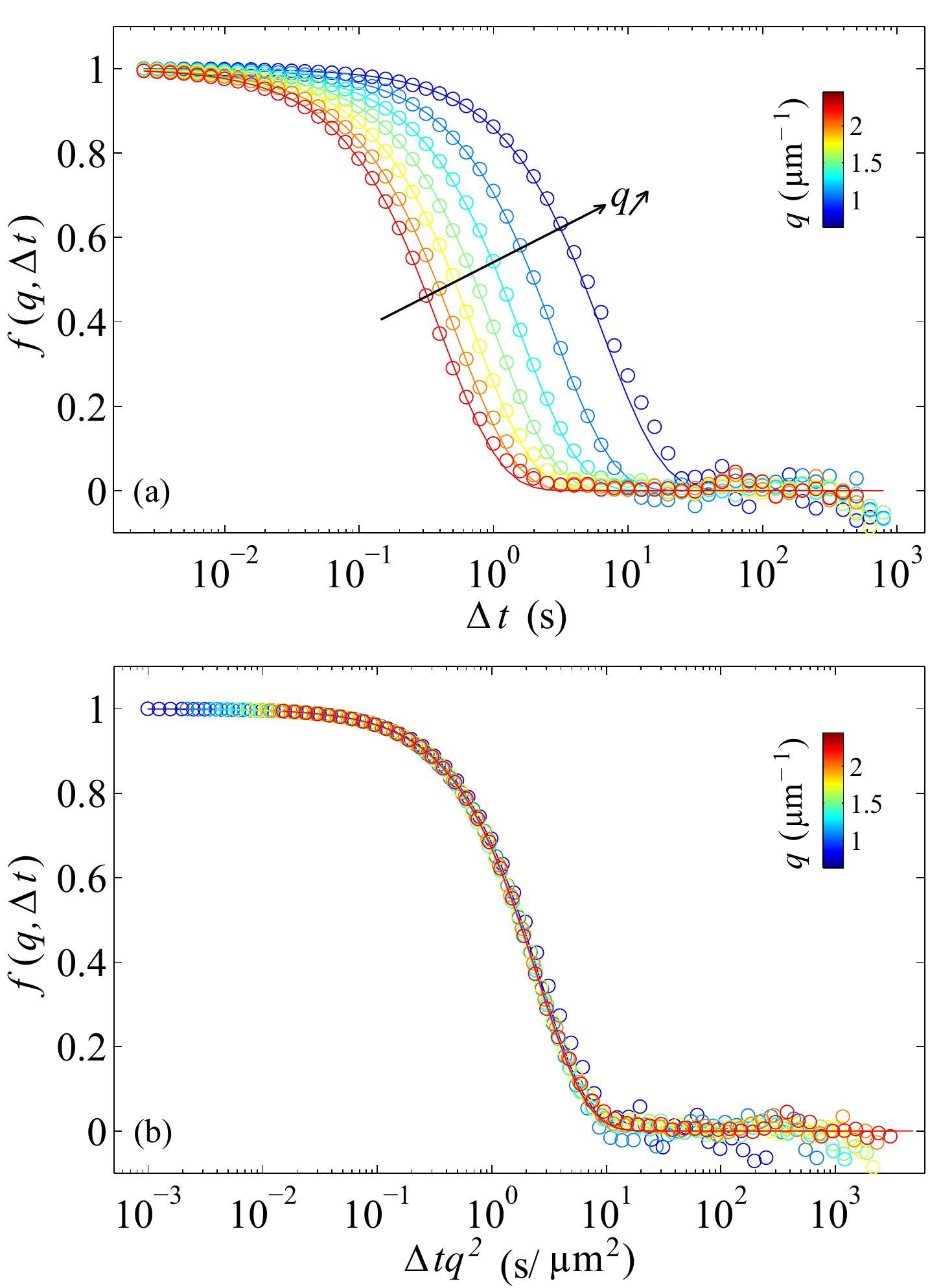}\\
	\caption{Autocorrelation function extracted from the DDM matrix $\mathcal{D}$ at various $q$ versus $\Delta t$ (a) and $\Delta tq^2$ (b). Lines are exponential fits to the data according to Eq.~\ref{eq:fdiff}.}
	\label{fig:ISF}
\end{figure}

Using strategy 1, from the short and long times values of $\mathcal{D}(q,\Delta t)$, we estimate $A(q)$ and $B(q)$ and isolate $f$, Fig.\ref{fig:ISF}.a. As shown in  Fig.\ref{fig:ISF}.b, $f$ collapse on a master curve when plotted as a function of $\Delta t q^2$. This scaling is compatible with a diffusive process and justified by Eq.~\ref{eq:fdiff}. 

Now we use the strategy 2. We incorporate the model in Eq. \eqref{eq:D2f} for $f$ in $\mathcal{D}$ and fit $\mathcal{D}$. The initial parameters for the fit are:
\begin{equation}
\left\{
\begin{array}{rcl}
A_0 &=& \mathcal{D} (q,\Delta t_\text{max}) - \mathcal{D} (q,\Delta t_\text{min}) \\
B_0 &=& \mathcal{D} (q,\Delta t_\text{min}) \\
\tau_\text{d} &=& \SI{1}{\second}
\end{array}
\right.
\end{equation}
where $\Delta t_\text{max}$ and $\Delta t_\text{min}$ are respectively the maximum and the minimum interval of time between two images.

\begin{figure}
	\includegraphics[width=\figwidth]{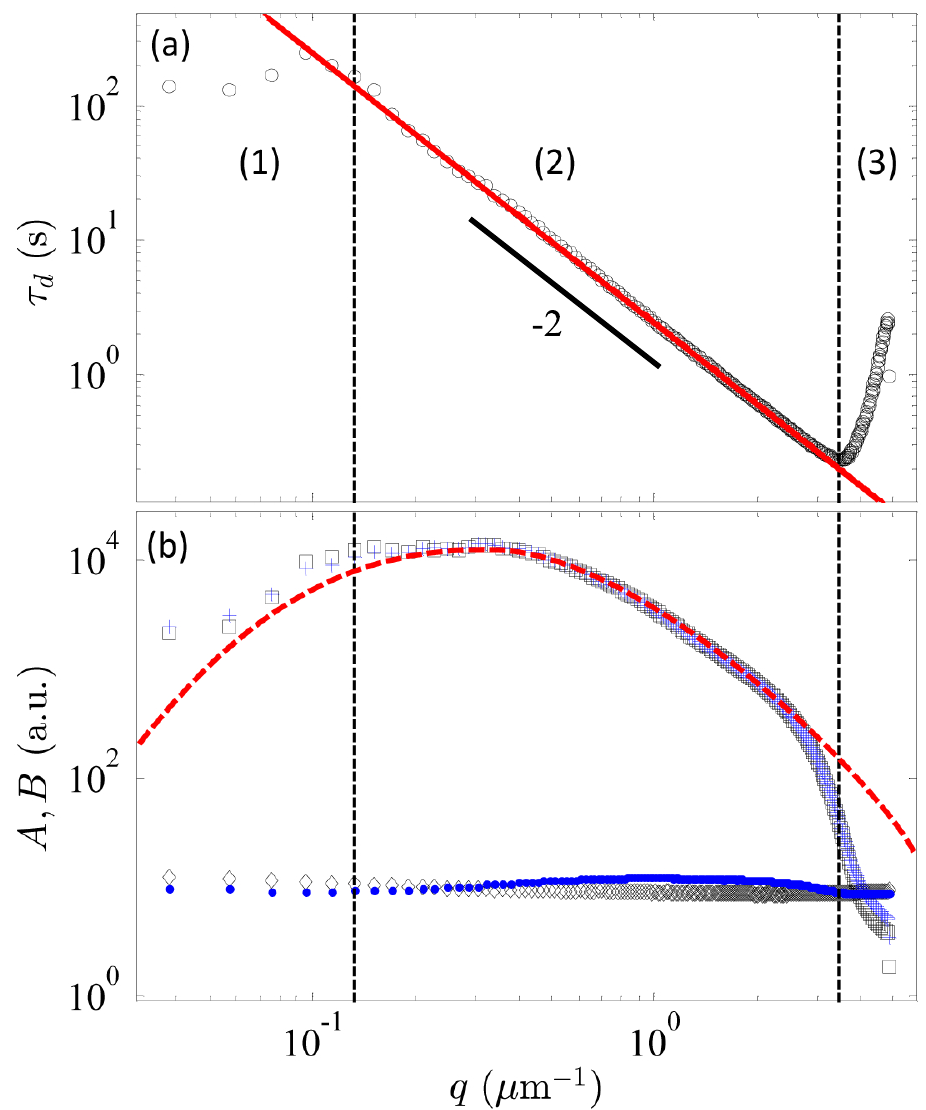}\\
	\caption{Characteristic Brownian diffusion time and the parameters $A(q)$, $B(q)$ versus $q$. (a) $\tau_\text{d}$ ($\circ$) and its fit (\textcolor{red}{\textbf{---}}) using Eq.\ref{eq:relax}. (b)  ($A$ (\textcolor{blue}{+}), $B$ (\textcolor{blue}{.})) measured with strategy 1  and ($A$ ($\Box$), $B$ ($\diamond$))  fitted with strategy 2. The red dashed line is a fit of $A(q)$ following the model in Ref\citep{3_DDM3D}. The vertical black dash lines delimite regime 1, 2 and 3. Fits are performed in regime 2.} 
	\label{fig:FitDiffColl}
\end{figure}

The fit results are displayed  as a function of $q$ in Fig.\ref{fig:FitDiffColl}. We observe 3 regimes:
\begin{itemize}
\item[\textbf{(1)}] Insufficient statistics. The radial average is performed on very few pixels (4 pixels for $q = q_\text{min}$ which correspond to the central cross of the Fourier transforms).  Also, at small $q$ the characteristic time is comparable to the duration of the experiment, $\SI{1000}{\second}$.
\item[\textbf{(2)}] Statistics are good and the signal-to-noise ratio is low enough.
\item[\textbf{(3)}] For  $q>\SI{3.5}{\per\micro\meter}$, $A(q)$ (signal) is too close to $B(q)$ (noise), to yield a consistent fit. This sets the spatial resolution of DDM to  $\frac{2\pi}{3.5} = \SI{1.8}{\micro\meter}$.
\end{itemize}

We fit $\tau_\text{d}(q)$ in regime (2) according to (Eq.~\ref{eq:relax}), which corresponds in logarithmic scale to a straight line of slope -2 and of intercept $-\log(D)$, Fig.\ref{fig:FitDiffColl}.a. Our measurements yield a diffusion coefficient of $D_\text{fit} = \SI{0.39}{\micro\meter\squared\per\second}$. We estimate the DDM experiment precision by repeating the exact same experiments 5 times. We obtain an average value $D_\text{fit}$ = 0.40 $\pm$ 0.02 $\mu$m$^2$/s with a relative precision of 5\%. The relative precision is comparable to the DLS technique\cite{Kayori2008}. Using the Stokes-Einstein formula with our experimental conditions and taking into account that the manufacturer stipulate a 10\% polydispersity on the colloidal radius we obtain $D_\text{E}$ = 0.42 $\pm$ 0.04 $\mu$m$^2$/s. The values $D_\text{E}$ and $D_\text{fit}$ are consistent.

There are many sources of uncertainties that may be reduced to increase the precision of the DDM results. First, statistics may be improved by increasing the number of particles in the field of view and the number of images in the time average. Second, DDM intrinsically deals with the static noise, thanks to the image subtraction process (Eq. \ref{eq:DI}), and the camera noise, $B$. To optimize the quality of the  measurement of $B$,one needs to measure fast enough so that $f$ remains almost completely correlated at short times. Otherwise $B$ is underestimated and consequently so is $\tau_\text{d}$.Third, $f$ needs to be properly normalized. One needs to measure long enough so that $f$ becomes completely decorrelated at long times otherwise $A+B$ is underestimated and consequently $\tau_\text{d}$ is overestimated. Experimentally this is easily checkable. One needs to observe for each $q$ two plateaus of the DDM matrix $\mathcal{D}$ at short and long lag times $\Delta t$, Fig.\ref{fig:D}e. The width of the range of $\Delta t$ is less critical for strategy 2 as all the parameters are fitted. Four, the choice of the region (2) where we fit the characteristic time also need to be properly estimated. Five, the results are model dependent. Ones need in fact to take the proper model for $f$. For example, the model in Eq.~\eqref{eq:fdiff}  considers that the particles are monodisperse. This is not exactly the case: the manufacturer stipulate a 10\% polydispersity. This could be refine by integrating a size polydispersity in the model for $f$. DDM is all the more precise that the experimentalists have some good knowledge about the system they study and have also first validated their procedure on a simple system like the one presented in this section.

 Fig.\ref{fig:FitDiffColl}.b shows the parameters $A(q)$ and $B(q)$. The noise level $B(q)$ seems constant for every $q$: the camera is adding a white noise to each images. With bright field, due to the depth of field of the objective in the $z$-direction, we are imaging a volume projected on $xy$ plane. The depth of field is an issue mostly on the large length scales (small $q$) due to the disappearance of particles from the $z$-field, which leads to underestimate the characteristic times\cite {4_Martinez20121637}. Taking into account the 3D nature of the experiment,  F. Giavazzi et al.\citep{3_DDM3D} have shown that it is possible to model $A(q)$.

In this section, we have verified and dissected the DDM procedure. We have shown that we obtain via the intermediate scattering function quantitative information on the 3D dynamics of a hundreds of colloidal particles simultaneously on length scales ranging from $\sim \SI{2}{\micro\meter}$ to $\sim \SI{200}{\micro\meter}$ and on time scales ranging from the millisecond to the minutes. This exact experiment can also be exploited in a different manner, thanks to the Stokes Einstein formula, Eq.\eqref{eq:se}. Provided  that we know $T$ and take for hydrodynamics radius of the particles $R= \SI{0.5}{\micro\meter}$, we can measure the viscosity of the solvant $\eta$. In our experiment, we find $\eta$= 1.07 mPas. Compared to a classical rheology experiment, DDM is actually better suited to measure small viscosity ($\sim$ mPas) of a solvent that we have on limited quantities ($\sim$ 50 $\mu L$). Finally, provided that this time we know $T$ take  $\eta$= 1.02 mPas, we can measure $R$ for particles size ranging from tens of nanometers to a few microns \cite{2_DDM}.  In our experiment, we find $R= \SI{0.53}{\micro\meter}$. Given the robustness and high throughput of DDM, DDM is appropriate for screening purposes.

\section{Bacteria and DDM}
\label{sec:BactSection}

\subsection{Motile Bacteria}

\begin{figure}
	\includegraphics[width=\figwidth]{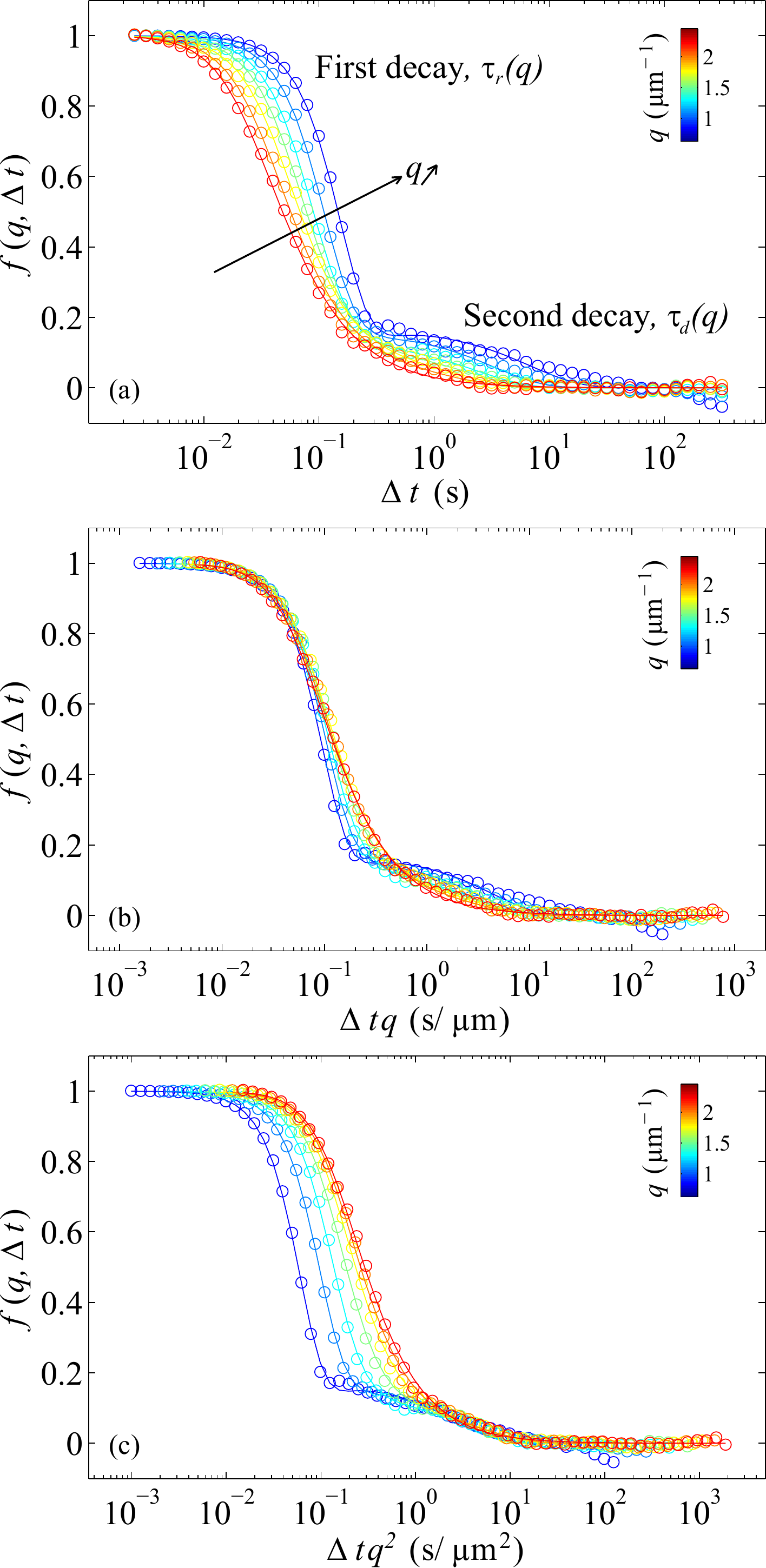}\\
	\caption{Autocorrelation function $f$ extracted from $\mathcal{D}$ at various $q$ versus $\Delta t$ (a), $\Delta tq$ (b) and $\Delta tq^2$ (c). Lines are fit to the data according to Eq.~\eqref{eq:fbact}.}
	\label{fig:ISFBacteria}
\end{figure}

\textit{The Salmonella Typhimurium SJW1103} dispersion has a more complex dynamics than colloidal dispersion. We used the same acquisition parameters and algorithm as for the colloids. We extracted $f$ from the DDM matrix using the strategy 1 where $A$ and $B$ are measure experimentally. Contrary to the colloidal case, $f(q, \Delta t)$ shows a two step decay which corresponds to two decorrelation mechanisms, Fig.\ref{fig:ISFBacteria}.a. The first decorrelation mechanism, for small $\Delta t$, is due to a ballistic motion of the bacteria: the first decays of $f$ collapses on a master curve as we scale $f$ with the abscissa  $\Delta t q$, see Fig.\ref{fig:ISFBacteria}.b. On the contrary, the second decorrelation mechanism observed at large $\Delta t$, is due to a diffusion process:  the second decays of $f$ collapse when plotted as a function of $\Delta t q^2$, see Fig.\ref{fig:ISFBacteria}.c.

Based on the scalings properties of the $f$, we turn to a model that takes into account the Brownian motion of bacteria due to $k_BT$, the mean velocity and the velocity distribution during the run, the fact that some of our bacteria are motile and some are not. Considering these new conditions, it can be shown that an adequate $f$ is \citep{1_BactMobil}:

\begin{align}
f(q, \Delta t) &= \exp\left(-\frac{\Delta t}{\tau_\text{d}}\right)\left[(1-\alpha) + \alpha \mathcal{P}(q, \Delta t)\right],\label{eq:fbact}\\
\mathcal{P}(q, \Delta t) &= \int_{0}^{\infty} P(v)\, \text{sinc}(\Delta t/\tau_\text{r}) \mathrm{d}v,\\
\text{with } &\tau_\text{d} = 1/(Dq^2) \text{ and } \tau_\text{r}=1/(q\overline{v}),\label{eq:taus}
\end{align}
with $\tau_\text{d}$ the characteristic Brownian diffusion time and $\tau_\text{r}$ the characteristic run time. The expression of $\tau_\text{d}$ and $\tau_\text{r}$ justify the scaling proposed in Fig. \ref{fig:ISFBacteria}.b and c.

$\alpha$ is the fraction of motile bacteria. $\text{sinc}(\Delta t/\tau_\text{r})$ is the $f$ of an isotropic population of swimmers at velocity $v$. The distribution of velocity $P(v)$ and the integral $\mathcal{P}(q, \Delta t)$ over $v$ take into account that all bacteria do not move at the same velocity. Following Wilson et al.\cite{1_BactMobil}, we chose a Schulz distribution for $P(v)$ which is peaked around the average velocity $\overline{v}$ and going to 0 when $v \rightarrow \infty$:
\begin{equation}
P(v) = \frac{v^Z}{Z!} \left(\frac{Z+1}{\overline{v}}\right)^{Z+1} \exp\left[-\frac{v(Z+1)}{\overline{v}}\right],
\end{equation}
with $Z$ is a parameter related to the standard deviation $\sigma$ of the velocity distribution $P(v)$:
\begin{equation}
Z = \left( \frac{\overline{v}}{\sigma} \right)^2 -1.
\label{eq:sig}
\end{equation}
The integral $\mathcal{P}$ can be formally calculated:
\begin{equation}
\mathcal{P}(q, \Delta t) = \frac{\sin\left(Z\tan^{-1}\theta\right)}{Z\theta\left(1+\theta^2\right)^{Z/2}}\text{, with }\theta = \frac{\Delta t}{\tau_r(Z+1)}. 
\end{equation}

\subsection{Results}
Using strategy 2, the fit of $\mathcal{D}(q, \Delta t)$ with Eq.~\eqref{eq:fbact} as model for $f$ requires 6 parameters. Even though the 2 decays of $f$ are well separated in time, we initialize the fit with values very close to the results so that the fit converges:
\begin{equation}
\left\{
\begin{array}{rcl}
A_0 &=& \mathcal{D} (q,\Delta t_\text{max}) - \mathcal{D} (q,\Delta t_\text{min}) \\
B_0 &=& \mathcal{D} (q,\Delta t_\text{min}) \\
\tau_\text{d0} &=& 1/(0.1 q^{2})\\
\tau_\text{r0} &=& 1/(10 q)\\
\alpha_0 &=& 0.5\\
Z_0 &=& 1
\end{array}
\right.
\end{equation}

\begin{figure}
	\includegraphics[width=\figwidth]{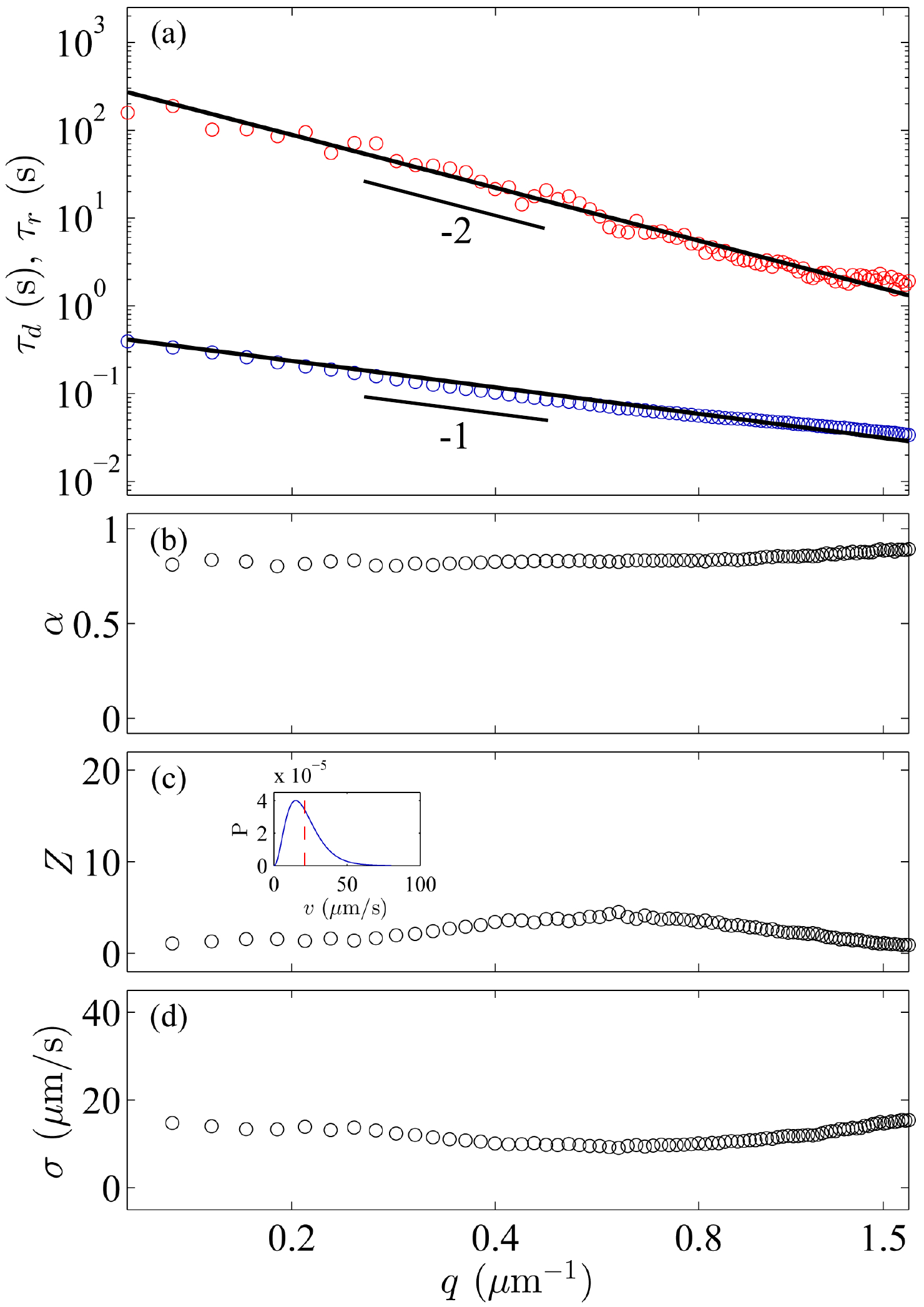}\\
	\caption{Fit parameters for the motile bacteria as a function of $q$  as extracted from the fit of the measurements of the DDM matrix $\mathcal{D}$. (a) characteristic time for the diffusion, $\tau_d$ (\textcolor{red}{$\circ$}), and balistic motion, $\tau_r$ (\textcolor{blue}{$\circ$}). (b) Fraction of motile bacteria $\alpha$. (c) $Z$. Inset: Shulz velocity distribution, $P$ for $<Z>$=2.4. (d) Standard deviation of the Shulz distribution $\sigma$ obtained from Eq. \eqref{eq:sig}.}
	\label{fig:FitParametersBacteria}
\end{figure}

The fit-output parameters are displayed as a function of $q$ in Fig. \ref{fig:FitParametersBacteria}. Their values are quite robust : repeating this experiment five consecutive times, yields a 6\% relative deviation. We limit the $q$ range to the regime (2) set by the colloid experiment where the statistics and the signal to noise are optimal. Fig. \ref{fig:FitParametersBacteria}a display the run and the diffusion time as the function of $q$ which are fitted in logarithmic scale by straight lines of respective slopes -1 (ballistic) and -2 (diffusion). The intercept of $\tau_\text{d}$ yields a diffusion coefficient of  $D = \SI{0.28}{\micro\meter\squared\per\second}$. The \textsc{Stokes-Einstein} relation considering spherical bacteria of diameter $\SI{1.5}{\micro\meter}$, $\eta = \SI{1}{\milli\pascal\second}$ and $T=\SI{293}{\kelvin}$, yields $D_\text{SE} = \SI{0.28}{\micro\meter\squared\per\second}$, a very good agreement even though we are not considering the real shape of bacteria. Fitting $\tau_\text{r}$ and using the mean value of $\sigma$ (Fig. \ref{fig:FitParametersBacteria}c-d) yields the mean velocity and the standard deviation of the bacteria in the ''run" state: $\overline{v}$ = 21.2 $\pm$ 11.3 $\mu$m/s. The \textit{Salmonella Typhimurium SJW1103}  are faster than \textit{E. Coli} ($\overline{v} \approx \SI{10}{\micro\meter\squared\per\second}$)\cite{4_Martinez20121637}. In Fig.~\ref{fig:FitParametersBacteria}b, we show the fraction of motile bacteria as a function of $q$. Our bacteria suspension displays a higher fraction of motile bacteria, $\alpha\approx$ 0.8 than \textit{E. Coli} ($\alpha\approx 0.6$)\cite{4_Martinez20121637}. This is why we chose this species rather than \textit{E. Coli}. DDM permits to measure the velocity along the trajectories of the bacteria on relevant length scales, up to $\sim 30 \mu$m, a clear advantage compared to the typical DLS experiment where the smallest $q$ only go down to $\sim$ 6 $\mu$m$^{-1}$, corresponding to $\sim$ 1 $\mu$m in real space, roughly size of a bacterium. We note that $\alpha$, $Z$ and $\sigma$ are expected to be constant as the statistical properties of the bacteria dispersion and should not change over the duration of the experiment nor with $q$. This is roughly what we observe. 


\section{Didactic considerations}
\label{sec:didac}
We had the chance to test this lab work on an unusually long format: 48h spread on 6 days. However, provided that the DDM code is already written~\cite{code_github}, and the acquisition parameters given to the students, the colloid part of this paper can be accomplished in a regular lab class of 8h. In this reduced format the students can make the sample, use the microscope and the camera, put in practice the concept of diffusion, build a Peclet number, understand the importance of the auto-correlation function and get Fourier Transform. A similar lab class on bacteria is more involved and probably to be kept for students who have already done the lab class on colloids.

An other option is a physics-inspired computer project~\cite{ajp2005spencer, ajp2014deutsh}. In this case, the students start from the movies in EPAPS and have to write the DDM code. Before any coding, we discuss with the students how to translate the principle of DDM into an algorithm, section \ref{sec:ddm}. To keep the code structured and readable for us we give them the signature of each function to code with predefined input and output. Students have to first produced a non optimized version of the algorithm to run on only 100 images. In this way the students can feel that unoptimized  calculations are heavy and unpractical for longer movies and can then optimise the algorithm as explained in section~\ref{sec:ddmalgo}. All along this process, we discuss the nitty-gritty of the Fourier transform, units conversion and the Nyquist–Shannon sampling theorem.

\section{Conclusion}
\label{sec:cl}
Differential Dynamic Microscopy (DDM) is a microscopy technique that probes the dynamics of a system of particles using a microscope, a camera and numerical computations. We exposed this technique in the well-known case of simple Brownian motion before applying it to the more complex case of the motile bacteria. We have shown how to extract physically relevant information from DDM based on scaling and how to obtain quantitative values such as the diffusion coefficient or the velocity.

DDM is based  on microscopy and can therefore benefit from  more advanced techniques than bright field imaging. For example, it is possible to use fluorescence microscopy to tell apart colloidal probes in a crowed medium\cite{Hendricks2015}, confocal microscopy\cite{Lu2012}, or polarized light microscopy\citep{20_reufer2012differential}. Apart from colloidal Brownian motion, DDM has also been used to characterize ellipsoidal particles\citep{20_reufer2012differential}, kinetics of phase separation\cite{Gao2015}, aggregation\cite{Ferri2011} and other species of bacteria, such as \textit{E. Coli} or \textit{C. Reinhardtii}\cite{4_Martinez20121637}. Even macroscopic systems, like human crowds shot from above, could be studied by DDM. In that sense, DDM opens much more possibilities than dynamic light scattering and can be used in various contexts both in the lab and in the classroom.

\section*{EPAPS}
\label{sec:epaps}
EPAPS are accessible as zip file at: \url{http://perso.ens-lyon.fr/thomas.gibaud/ddm} and  contains 4 movies and a note describing each movies. Each movies are obtained with bright field microscopy and are composed of 4000 images of size = $(\SI{512}{px})^2$.  $\SI{1}{px}$ represent $\SI{0.645}{\micro\meter}$. Movie 1 and 2 capture the dynamics of the colloidal dispersion at $f_{acq}$=400 and 4 Hz. Movie 3 and 4 capture the dynamics of the \textit{Salmonella} dispersion at $f_{acq}$=400 and 4 Hz.

We have attributed a DOI \href{http://dx.doi.org/10.5281/zenodo.30559}{10.5281/zenodo.30559}\cite{code_github} to the codes which are accessible at GitHub.

\begin{acknowledgments}
The authors would like to thank the students who endured the initial versions of this lab work: Alicia Damm, Florine Dubourg, D.G. himself, Paul Haddad, Gabriel Rigon and Sylvie Sue. 
This work was supported by the ENS Lyon, the Agence Nationale de la Recherche fran\c{c}aise (ANR-11-PDOC-027) and ML thanks the Region Rh\^one Alpes and
the Programme d'Avenir Lyon - Saint Etienne (PALSE NoGELPo) for postoctoral grant.
\end{acknowledgments}

\bibliographystyle{apsrev4-1}
\bibliography{DGermain_2015bbib}

\end{document}